\documentclass[aps,prl,reprint]{revtex4-1}

\usepackage{graphicx}
\usepackage{dcolumn}
\usepackage{bm}
\usepackage[utf8]{inputenc}
\usepackage[T1]{fontenc}
\usepackage{mathptmx}
\usepackage{amsmath}
\usepackage{amssymb}
\usepackage{physics}
\usepackage{booktabs}
\usepackage[english]{babel}

\begin{document}

\title{Biermann-battery driven magnetized collisionless shock precursors in laser produced plasmas}

\author{T. M. Johnson}
\email{tmarkj@mit.edu}
\affiliation{
    Plasma Science and Fusion Center, Massachusetts Institute of Technology,
    Cambridge, Massachusetts 02139, USA 
}

\author{G. D. Sutcliffe}
\affiliation{
    Plasma Science and Fusion Center, Massachusetts Institute of Technology,
    Cambridge, Massachusetts 02139, USA 
}

\author{J. A. Pearcy}
\affiliation{
    Plasma Science and Fusion Center, Massachusetts Institute of Technology,
    Cambridge, Massachusetts 02139, USA 
}

\author{A. Birkel}
\affiliation{
    Plasma Science and Fusion Center, Massachusetts Institute of Technology,
    Cambridge, Massachusetts 02139, USA 
}

\author{G. Rigon}
\affiliation{
    Plasma Science and Fusion Center, Massachusetts Institute of Technology,
    Cambridge, Massachusetts 02139, USA 
}

\author{N. V. Kabadi}
\affiliation{
    Plasma Science and Fusion Center, Massachusetts Institute of Technology,
    Cambridge, Massachusetts 02139, USA 
}

\author{B. Lahmann}
\affiliation{
    Plasma Science and Fusion Center, Massachusetts Institute of Technology,
    Cambridge, Massachusetts 02139, USA 
}

\author{P. J. Adrian}
\affiliation{
    Plasma Science and Fusion Center, Massachusetts Institute of Technology,
    Cambridge, Massachusetts 02139, USA 
}

\author{B. L. Reichelt}
\affiliation{
    Plasma Science and Fusion Center, Massachusetts Institute of Technology,
    Cambridge, Massachusetts 02139, USA 
}

\author{J. H. Kunimune}
\affiliation{
    Plasma Science and Fusion Center, Massachusetts Institute of Technology,
    Cambridge, Massachusetts 02139, USA 
}

\author{S. G. Dannhoff}
\affiliation{
    Plasma Science and Fusion Center, Massachusetts Institute of Technology,
    Cambridge, Massachusetts 02139, USA 
}

\author{M. Cufari}
\affiliation{
    Plasma Science and Fusion Center, Massachusetts Institute of Technology,
    Cambridge, Massachusetts 02139, USA 
}

\author{F. Tsung}
\affiliation{
    Department of Physics and Astronomy, University of California Los Angeles, Los Angeles, California 90095, USA
}

\author{H. Chen}
\affiliation{
    Lawrence Livermore National Laboratory, Livermore, California 94550, USA
}

\author{J. Katz}
\affiliation{
    Laboratory for Laser Energetics, University of Rochester, Rochester, New York 14623, USA
}

\author{V. T. Tikhonchuk}
\affiliation{
    Centre Lasers Intenses et Applications, University of Bordeaux, CNRS, CEA, 33405 Talence, France
}
\affiliation{
    The Extreme Light Infrastructure ERIC, ELI Beamlines Facility, 252 41 Doln\'{i} B\u{r}e\u{z}any, Czech Republic
}

\author{C. K. Li}
\email{ckli@mit.edu}

\affiliation{
    Plasma Science and Fusion Center, Massachusetts Institute of Technology,
    Cambridge, Massachusetts 02139, USA 
}

\date{\today}

\begin{abstract}
    This letter reports the first complete observation of magnetized collisionless shock precursors
    formed through the compression of Biermann-battery magnetic fields in laser produced plasmas.
    At OMEGA, lasers produce a supersonic CH plasma flow which is magnetized with Biermann-battery
    magnetic fields.
    The plasma flow collides with an unmagnetized hydrogen gas jet plasma to create a magnetized shock precursor.
    The situation where the flowing plasma carries the magnetic field is similar to the Venusian bow shock.
    Imaging 2$\omega$ Thomson scattering confirms that the interaction is 
    collisionless and shows density and temperature jumps.
    Proton radiographs have regions of strong deflections and FLASH 
    magnetohydrodynamic (MHD) simulations show the presence of Biermann fields 
    in the Thomson scattering region.
    Electrons are accelerated to energies of up to 100 keV in a power-law spectrum. 
    OSIRIS particle-in-cell (PIC) simulations, initialized with measured parameters, 
    show the formation of a magnetized shock precursor and corroborate the 
    experimental observables.
\end{abstract}

\maketitle


Collisionless shocks are very common in astrophysical systems.
Counter-streaming plasmas, ranging from Earth's magnetosphere\cite{Mozer1981-ro} 
to relativistic astrophysical jets\cite{Medvedev1999-xm}, often form shocks which dissipate energy.
Charged particles can be accelerated to high energies inside shocks\cite{Ackermann2013-gb}.
Magnetized shocks are one type of collisionless shock\cite{Sakawa2016-jz}.
They form when a dynamically significant magnetic field is present 
in a system of counter-streaming plasmas and are very common in astrophysics\cite{Treumann2009-pc}.
The majority of planetary bow shocks, an example of a magnetized collisionless shock,
form from the interaction between the weakly magnetized solar wind and a strongly magnetized 
planetary ionosphere\cite{Balogh2013-ps}.
Venus, however, has no magnetic field making the solar wind 
field responsible for its bow shock\cite{Luhmann1986-fw}.
This is one astrophysical situation where the flowing plasma carries the magnetic field 
responsible for shock formation.

This letter reports the first complete observation of a Biermann-battery driven 
magnetized collisionless shock precursor\cite{Biermann1951-oz}.
There are no externally imposed magnetic fields.
Instead, Biermann-battery fields, generated during the laser drive,
are frozen into the plasma flow.
These fields are compressed in the collision between the plasma flow
and gas jet plasmas.
The magnetic field strength is enhanced, causing gas jet ions to be deflected 
and a magnetized shock precursor to be formed. 
Since the flowing plasma carries the magnetic field, 
the presented experiment is similar to the interaction between the solar wind and 
the Venusian ionosphere\cite{Luhmann1986-fw}.
The origin of nightside aurorae on Venus is currently unknown\cite{Kovac2022-lx}.
Charged particles accelerated by the bow shock could be responsible.

While magnetized shocks relevant to planetary bow shocks have been studied in the 
laboratory, all experiments have focused on the case where the stationary plasma 
contains the magnetic field\cite{Schaeffer2017-vy, Schaeffer2019-am, Yao2021, Suzuki-Vidal2015-fw, Levesque2022-mg}.
The experiment presented here demonstrates a platform to study Venus's particular 
configuration where the flowing plasma carries the magnetic field.
Such configurations have been studied previously, but the results lacked 
direct evidence of the magnetic field\cite{Yamazaki2022-tc, Umeda2019-dg}.
Other experiments with a plasma flow colliding with a gas bag produced an 
electromagnetic shock structure, but the gas bag shell
played a significant role in the overall physics of the experiment\cite{Li2019-qt}.
Additionally, the results of this experiment show that Biermann-battery generated 
magnetic fields can be strong enough to dominate the physics of laser-produced 
high-energy-density plasmas.
This conclusion differs from studies of electromagnetic shocks 
with planar foils which found that Biermann-battery magnetic fields were 
not dynamically important to the overall interaction\cite{Huntington2015-ak}.


\begin{figure}[h!]
    \begin{center}
        \includegraphics[width=.45\textwidth]{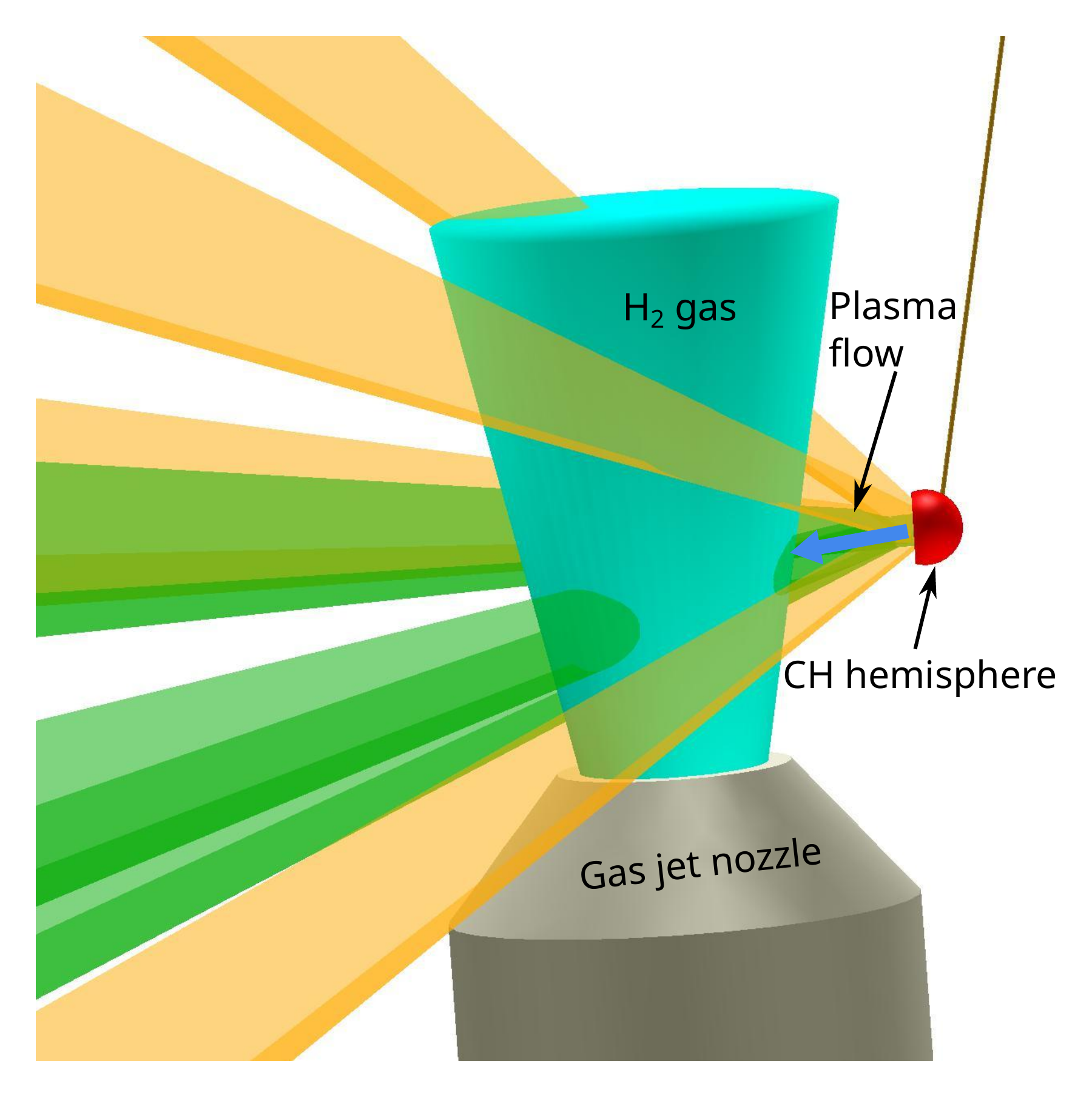}
    \end{center}
    \caption{
        Side view of the OMEGA experimental configuration showing the gas jet, the hydrogen gas volume,
        and the CH hemispherical target. 
    }
    \label{fig:exp_config}
\end{figure}

A schematic of the OMEGA experiment geometry is shown in Fig.\ \ref{fig:exp_config}.
The gas jet produces a volume of hydrogen gas.
Seven 351 nm laser beams each deliver 500 J of energy to a CH hemisphere 
in a 1 ns square pulse to produce a plasma flow.
The gas jet volume is ionized prior to the arrival of the plasma flow.
The interaction between the gas jet plasma and the plasma flow
is diagnosed with three diagnostics.
Imaging 2$\omega$ Thomson scattering measures the density,
temperature, and velocity profiles at different times.
Proton radiography, using a D$^3$He backlighter, records particle deflections from 
electromagnetic fields.
Electron spectroscopy measures the acceleration of electrons.


\begin{figure*}[t!]
    \centering
    \includegraphics[width=1.0\textwidth]{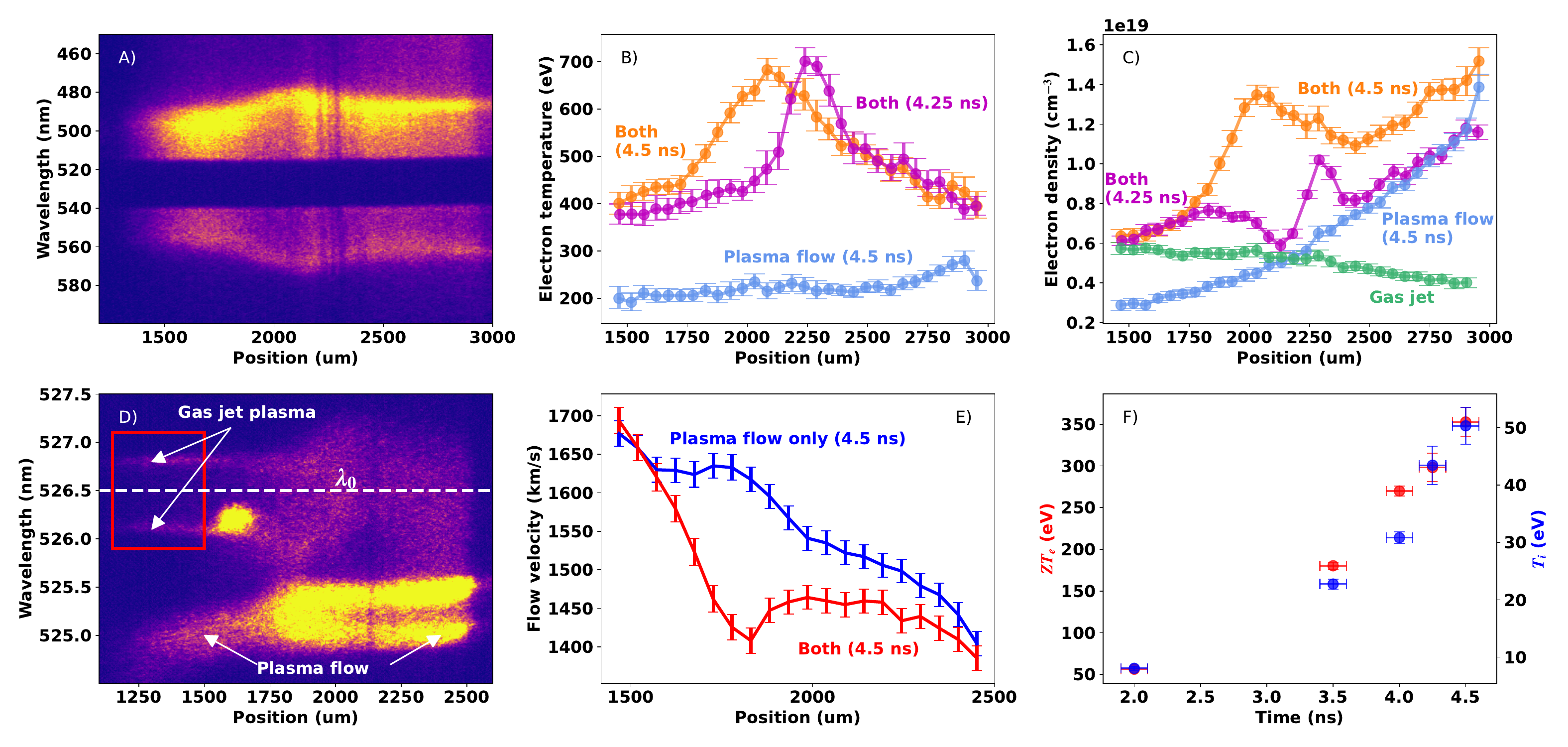}
    \caption{
        A) Thomson EPW spectrum at 4.5 ns for shots with both 
        the gas jet and the plasma flow. 
        Distance is measured away from TCC along the plasma flow axis. 
        B) Electron temperature profiles for the interaction at different times 
        compared to a plasma flow only shot.
        C) Electron density profiles for the interaction compared to the plasma flow and gas jet
        profiles.
        D) Thomson IAW spectrum at 4.5 ns showing the gas jet plasma spectrum around the probe wavelength
        and the blue shifted plasma flow spectrum.
        E) Flow velocity profile at 4.5 ns for shots with and without the gas jet.
        F) Time resolved and spatially averaged (from red box region in D) $ZT_e$ and $T_i$ of the gas jet plasma
        ahead of the shock.
    }
    \label{fig:thomson_fig}
\end{figure*}

Imaging $2\omega$ Thomson scattering collects spatially resolved 
electron plasma wave (EPW) and ion acoustic wave (IAW) data
at different times\cite{Sheffield2010-uh, Katz2013-tc, Katz2012-do}.
The Thomson probe beam points directly down
the plasma flow axis and focuses to a region 2 mm from target chamber center (TCC).
The spatial field of view is about 1.5 mm long in the direction of the probe beam
with the Thomson scattering $\vb{k}$ vector 59.885 degrees off axis.
Fig.\ \ref{fig:thomson_fig} shows results from EPW and IAW measurements.
All times reference the start of the laser drive on the hemispherical target.

Fig.\ \ref{fig:thomson_fig} B) and C) show enhancements of the electron temperature and density
due to the interaction between the gas jet and the plasma flow.
Comparing the location of the density jump across different times shows that the feature
has a velocity of $1000 \pm 200$  km/s.
The width of the density peak at 4.5 ns is $\sim300$ $\mu$m 
(compared to ion skin-depth, $\sim100$ $\mu$m, and Larmor radius, $\sim500$ $\mu$m).

Fig.\ \ref{fig:thomson_fig} D) shows the raw IAW spectrum which 
contains two plasma species.
The spectrum centered around the probe wavelength is the gas jet 
plasma since it has no appreciable flow velocity.
Its narrow peaks indicate that $ZT_e/T_i$ is large.
Blue shifted from the probe is the plasma flow spectrum.
The flow velocity is close to the flow velocity of the plasma flow without the gas jet plasma,
but with a localized velocity dip seen in Fig.\ \ref{fig:thomson_fig} E).

Flow velocity measurements of the plasma flow confirm low ion-ion collisionality.
Fig.\ \ref{fig:thomson_fig} E) shows the flow velocity profiles with and without the gas jet.
For the plasma flow only, the velocity exceeds 1500 km/s.
The velocity increases farther away from the CH hemisphere target\cite{Mora2003-hi, Ross2012-POP}.
With flow velocity and density measurements, the interspecies ion-ion collision mean-free-path can be calculated:

\begin{equation}
    \label{eq:1}
    \lambda_\text{mfp} = \frac{(4\pi\epsilon_0)^2}{n_2Z_1^2 Z_2^2e^4}\frac{m_1m_2v_1^4}{8\pi\log\Lambda}
\end{equation}

\noindent
where indices $1$ ($2$) refers to the plasma flow (gas jet), $n$ is the ion density,
$Z$ is the charge state, $m$ is the ion mass, and $\log \Lambda$ is the 
Coulomb logarithm\cite{Park2012-db}.
For the experiment, the plasma flow carbon ion mean-free-path is about 7 cm 
which is much larger than the system size and the density peak width.
Therefore, the interaction between the plasma flow ions and the gas jet ions is collisionless.

Fig.\ \ref{fig:thomson_fig} F) shows time resolved measurements of $ZT_e$ and $T_i$ for
the gas jet plasma in front of the density jump.
The red box in Fig.\ \ref{fig:thomson_fig} D) shows the region where $ZT_e$ and $T_i$ are
measured.
If $Z$ is assumed to be one (gas is hydrogen), 
the $ZT_e$ values agree with the EPW measured $T_e$ values 
in front of the temperature peak.
This heating is caused by electron-ion collisional 
friction\cite{Ross2012-POP}.
Such heating is spatially uniform which does not 
explain the observed the localized temperature jump.

The density jumps by $2.53\pm0.15$ times and the temperature jumps by
$1.94\pm0.12$ times (at 4.5 ns, the gas jet has been heated to about 350 eV).
These jumps are measured with respect to the gas jet.
The measured jumps do not match the Rankine-Hugoniot conditions for the 
sonic Mach number of $\sim4$.
This is due to the interaction being only a shock precursor\cite{Schaeffer2020}.  
Not enough time has elapsed for the shock to be fully formed.
At the probed time, the shock is still developing as seen in Fig.\ \ref{fig:thomson_fig} C)
where the density peak is increasing with time.

\begin{figure}[h]
    \centering
    \includegraphics[width=.45\textwidth]{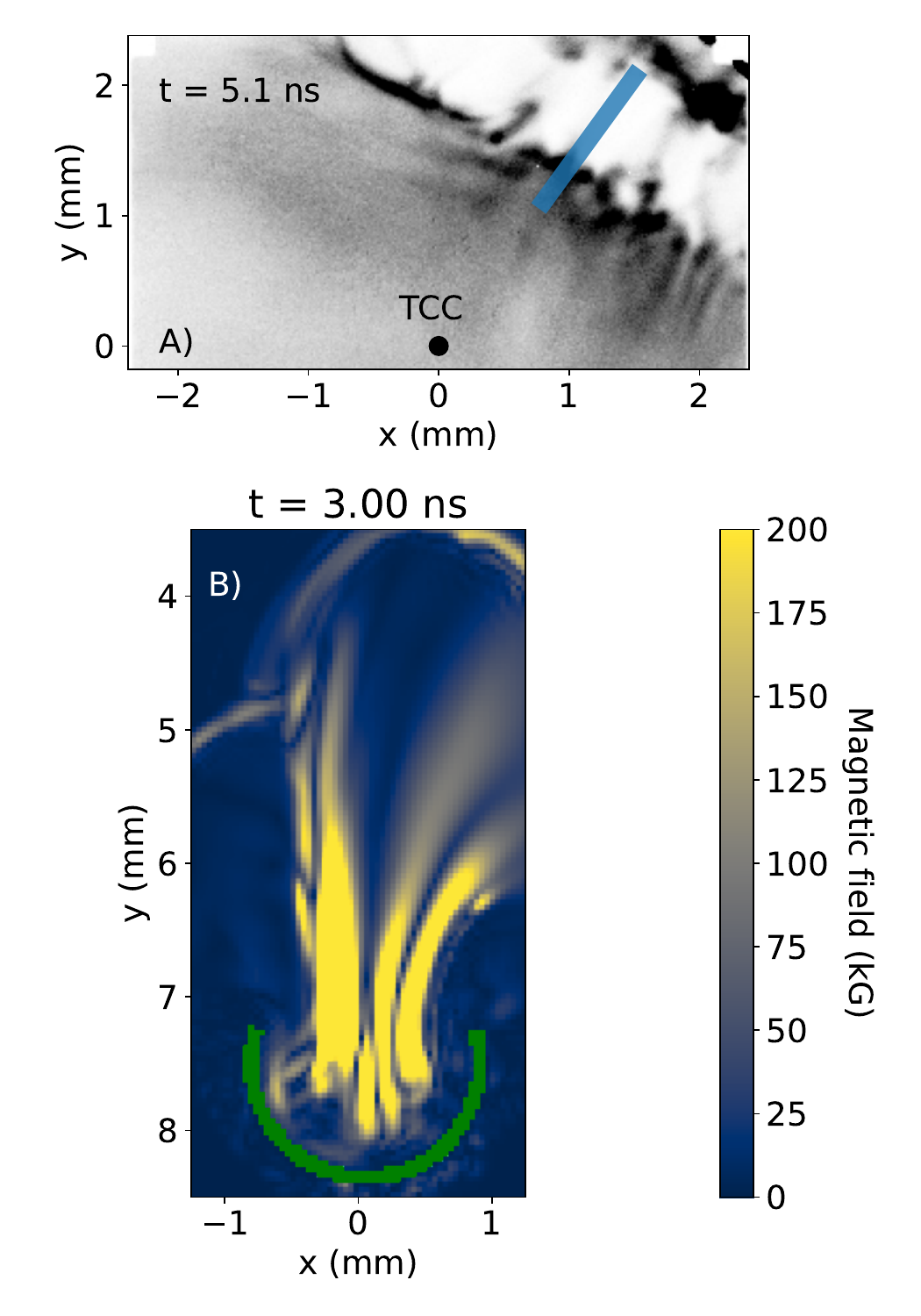}
    \caption{
        A) Radiograph from 3 MeV protons taken at 5.1 ns showing a sharp deflection region
        of strong field. 
        The blue region shows the Thomson scattering volume projected into the radiograph field-of-view.
        B) 2D slice from the 3D FLASH simulation at 3.0 ns showing the magnitude of the 
        magnetic field perpendicular to the plasma flow.
        Note the region of strong magnetic field on the plasma flow axis.
        The hemisphere target is shown in green.
        The $y$ axis references the distance from TCC.
    }
    \label{fig:PRAD_FLASH}
\end{figure}

Proton radiography images the electromagnetic fields from the plasma flow gas jet interaction
using 3 MeV and 15 MeV protons\cite{Li2006-ng, Li2008-sv}.
Fig.\ \ref{fig:PRAD_FLASH} A) shows a resulting 3 MeV radiograph.
The radiograph has a region of strong deflection 
co-located with the Thomson scattering region.
There are no filamentary structures at the probed time meaning that the deflections
cannot be from the Weibel instability or other plasma 
instabilities\cite{Huntington2015-ak}.
The only source of fields are Biermann-battery fields from the laser drive.
Deflections in the radiographs come from magnetic fields since electric fields  
are ruled out.

A set of 3D Cartesian FLASH ideal MHD simulations with the Biermann-battery term 
model the plasma flow Biermann fields before 
the collision with the gas jet\cite{Fryxell2000-zn, Tzeferacos2015-jj}.
These simulations have the same laser conditions and geometry as in the experiment 
with the target shifted a realistic 50 $\mu$m in the $x$-direction.
The simulation results in Fig.\ \ref{fig:PRAD_FLASH} B) show the magnetic field topology.
Biermann fields are present along the plasma flow axis and therefore are present 
in the Thomson scattering volume, located right outside the simulation domain.
 

\begin{figure}
    \centering
    \includegraphics[width=.45\textwidth]{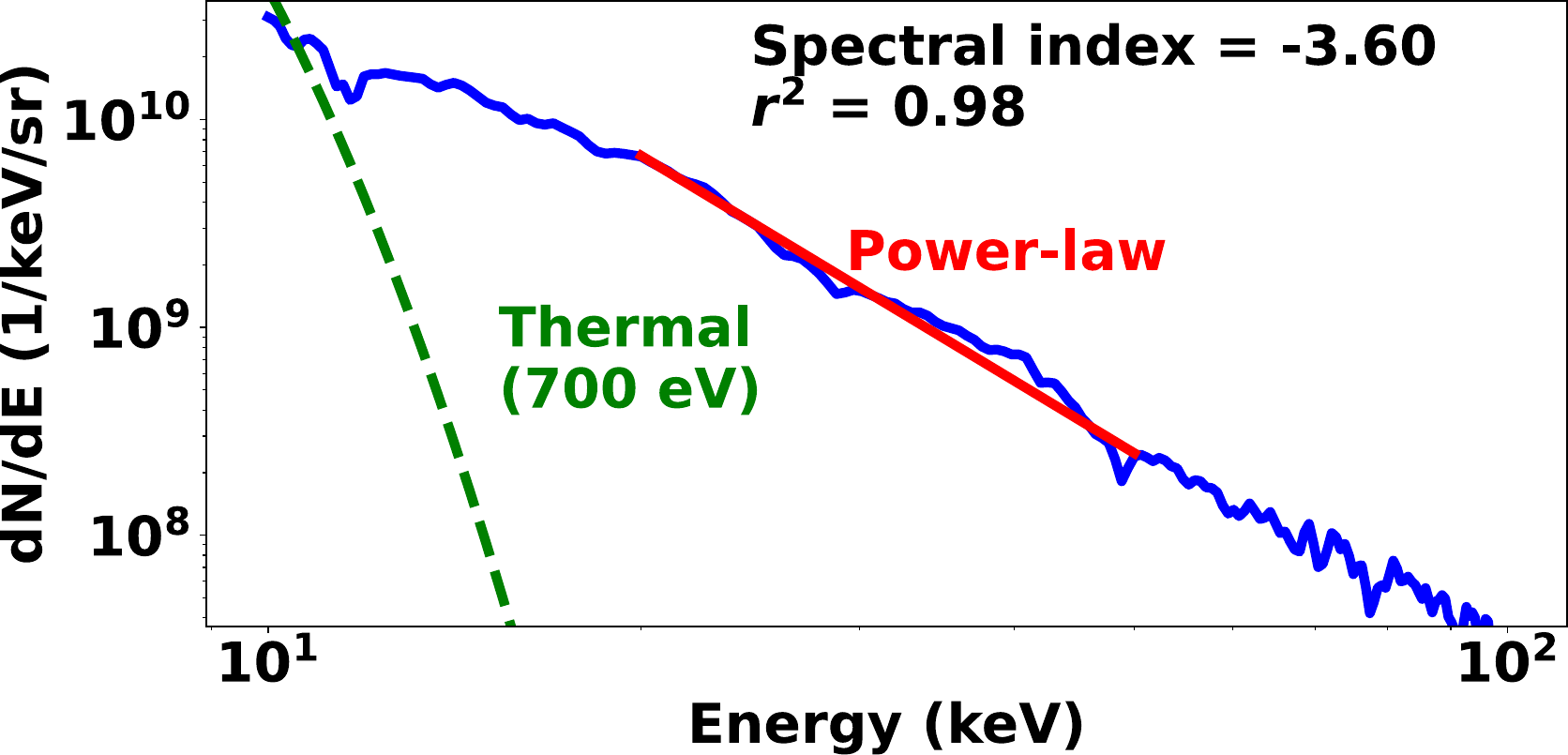}
    \caption{
        Log-log plot of the net electron spectrum.
        A fixed temperature Maxwellian is fit to the low energy emphasizing the 
        non-thermal nature of the spectrum.
        A power-law is fit to the high energy portion of the spectrum.
    }
    \label{fig:EPPS_fig}
\end{figure}

Electron spectroscopy measurements, using the Electron Positron Proton Spectrometer (EPPS) 
diagnostic\cite{Chen2008-vq, Chen2008-xf},
show the acceleration of electrons into a high energy power-law spectrum.
To find the net acceleration spectrum, shots with and without the gas jet are compared.
The results of this analysis are shown in Fig.\ \ref{fig:EPPS_fig}.
A Maxwellian with the maximum measured electron temperature
is fit to the low energy part of the spectrum to emphasize the high energy tail.
A power-law is fit to the high energy non-thermal part of the spectrum yielding a 
spectral index of -3.6 and giving clear evidence of electron acceleration.
The quality of the fit is confirmed through a simple $r^2$ analysis.
Stimulated Raman scattering from the laser passing through the gas jet is ruled out 
as a source of fast electrons.

\begin{figure*}
    \centering
    \includegraphics[width=1.0\textwidth]{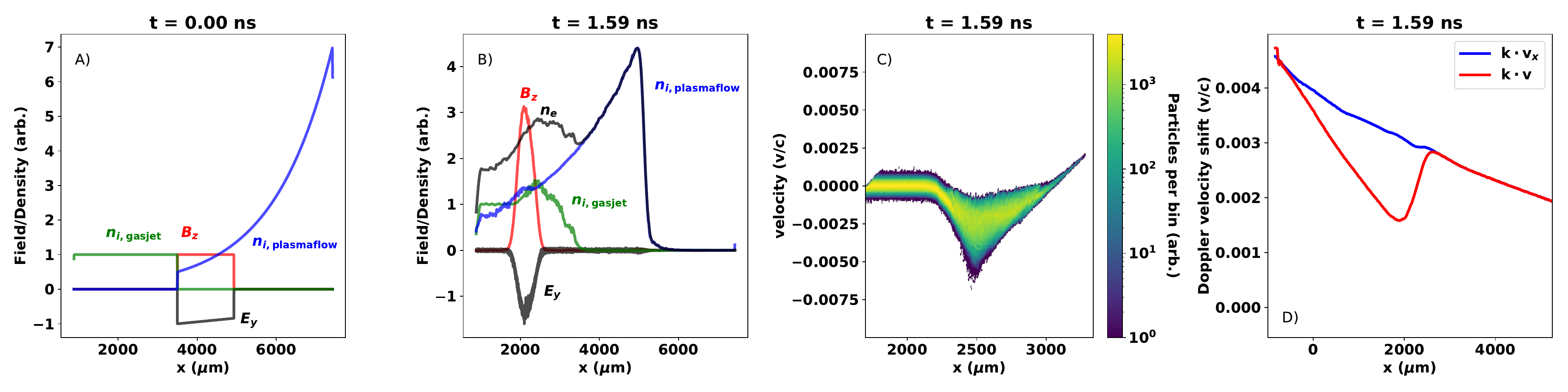}
    \caption{
        A) Initial simulation EM field and density profiles.
        B) Simulation profiles at $t = 1.59$ ns.
        C) Gas jet ion $x$-velocity $x$-position phase space plot at $t = 1.59$ ns.
        D) Thomson scattering-like Doppler shift of the plasma flow ions from deflection 
        in the magnetic field.
        The localized dip is consistent with Fig.\ \ref{fig:thomson_fig} E).
        In all plots, the $x$ axis references distance from TCC.
    }
    \label{fig:PIC}
\end{figure*}

Particle-in-cell (PIC) simulations are performed to study the kinetic aspects of the interaction.
OSIRIS 1D3V PIC simulations show the formation of a magnetized shock precursor for 
experimentally relevant conditions\cite{Fonseca2002-sq}.
The PIC simulations span 6000 $\mu$m of space, have a spatial resolution of 0.034 $\mu$m, and have 
1000 particles per cell and realistic mass ratios.
Fig.\ \ref{fig:PIC} A) shows the initial conditions of the simulation with a 
uniform density profile on the left serving as the gas jet plasma and a self-similar
density profile on the right serving as the plasma flow.
A region of the plasma flow has a uniform magnetic field of 75 kG with an associated
induction electric field to model the Biermann-battery fields.
The left (gas jet) plasma is stationary while the right (plasma flow) plasma flows 
into it with a velocity profile similar to Fig.\ \ref{fig:thomson_fig} E).

The simulation captures essential features of magnetized shock precursor formation.
Since the ions are collisionless, the plasma flow ions pass through the interface 
resulting in an increase in the total density.
The magnetic field in the plasma flow reflects the gas jet electrons meaning that 
the plasma flow electrons alone neutralize the ion charge, causing an increase
in the plasma flow electron density.
Since the magnetic field is frozen into the plasma flow (magnetic Reynolds number $\sim 380$), the increase in the 
plasma flow electron density also increases the magnetic field strength. 
Magnetic flux conservation requires that the magnetic field peak propagates forward
slower than the initial plasma flow velocity.
Simply put, the system forms a magnetized piston immediately after the collision\cite{Moreno2020-ox}.
Fig.\ \ref{fig:PIC} B) shows the profiles after the simulation has evolved.
The magnetic field is strong enough  to start deflecting the gas jet ions (gyro-radius $\sim500$ $\mu$m), 
seen in Fig.\ \ref{fig:PIC} C), increasing the ion density and moving the density 
peak away from the interface at a velocity of 950 km/s.
Since the plasma flow ions have a larger charge to mass ratio compared to the gas jet ions, they are stiffer 
to deflection causing the plasma flow density to be unaffected by the magnetic field.

The electric field is enhanced less than the magnetic field,
resulting in a net Lorentz force on the plasma flow ions in the magnetic field region.
As the plasma flow ions spend time in this region, their velocity in 
the $x$-direction is roughly unchanged, but they accrue a non-negligible 
velocity in the $y$-direction.
The Thomson IAW Doppler shift is sensitive to $\vb{k} \cdot \vb{v}$. 
In the experimental geometry, the angle between the probe $\vb{k}$
vector and the plasma flow axis is about 60 degrees.
Therefore, the Doppler shift is sensitive to the velocity in the $y$-direction:

\begin{equation}
    \label{eq:kdotv}
    \vb{k} \cdot \vb{v} = (k_i - k_s\cos(\theta_s))v_x - k_s\sin(\theta_s)v_y, 
\end{equation}

\noindent
where $k_i$ and $k_s$ are the magnitudes
of the probe and collected light wavevectors (assumed to be equal) 
and $\theta_s$ is the Thomson scattering angle.
Fig.\ \ref{fig:PIC} D) shows the Doppler shift with a localized dip due 
to the deflection of the plasma flow ions to the magnetic field which is consistent 
with the Thomson IAW velocity measurements shown in Fig.\ \ref{fig:thomson_fig} E).
Note that the brightness of the Thomson IAW features increases with $Z$ and density\cite{Sheffield2010-uh}.
While the gas jet ions are deflected, their Doppler shifted spectrum is outshone by the plasma flow ions.

The PIC simulations results match the experimental measurements well.
The velocities of the magnetic field and the density jump are measured in the simulation to be 870 km/s
and 950 km/s respectively, which is comparable to the experimentally measured 
velocity of $1000\pm200$ km/s.
The simulated density jump feature has a similar shape and peak value compared to the EPW measurements.
Deflections of the plasma flow ions produce similar Doppler shifts 
as the Thomson IAW data.


The formation process of the shock precursor is as follows.
Lasers illuminate the CH hemispherical target and generate strong $\sim$MG-scale 
Biermann-battery magnetic fields.
These fields are frozen into the plasma flow since the magnetic Reynolds number 
is large\cite{drake_book}.
The plasma flow expands as it travels, reducing the field strength.
A magnetic piston forms when the plasma flow and gas jet interpenetrate, 
enhancing the magnetic field strength.
Gas jet ions see the magnetic field and get deflected causing
the density jump to move forward.
The presence of the reflected upstream ions satisfies the 
criteria for a magnetized shock precursor\cite{Schaeffer2020}.
The resulting magnetized precursor has an Alfv\'en Mach number of $M_A\sim14$
making it supercritical\cite{Treumann2009-pc}.

The observed electron acceleration is not seen in the PIC simulations, 
likely due to limitations of the 1D simulation.
Given the enhanced magnetic field strength of $\sim200$ kG, 
different shock acceleration mechanisms can be considered.
The electron gyro-radius is too small for diffusive shock acceleration\cite{Bell1978-hu}
and the Alfv\'en Mach number is too small for shock surfing acceleration\cite{Matsumoto2012-nh}.
The last mechanism left is shock drift acceleration (SDA) where
electrons traveling along the magnetic field ramp are accelerated
by the induction electric field\cite{Wu1984-nk}.
The observed acceleration is therefore plausibly attributed to SDA.

The observed shock precursor differs from previous experiments
since it moves slower than the initial flow velocity\cite{Schaeffer2017-vy, Schaeffer2019-am}.
Viewed from the center-of-mass frame, the shock precursor 
is moving backwards towards the CH target.
If fully formed, the shock would be the reverse shock.
The counter-streaming of the plasma flow upstream would, if given enough time and energy, form a forward moving
electromagnetic shock via the beam-Weibel instability\cite{Moreno2020-ox}.

In summary, this letter details and explains the first complete observation of 
magnetized collisionless shock precursors in laser-driven plasmas without 
externally imposed magnetic fields.
The experiment offers a laboratory example of a situation similar to the formation 
of Venus's bow shock.
The observed electron acceleration could be relevant to the unknown origin of the nightside aurorae on Venus.

This work was supported, in part, by the U.S. Department of Energy NNSA MIT Center-of-Excellence under Contract No.
DE-NA0003868, by the National Laser Users Facility under Contract No. DE-NA0003938, and by the NNSA HEDLP program
under Contract No. DE-NA0004129.
Some of the simulations presented in this paper were performed on the MIT-PSFC partition of the Engaging cluster at the MGHPCC facility 
(www.mghpcc.org) which was funded by DoE grant number DE-FG02-91-ER54109.
The software used in this work was developed in part by the DOE NNSA and DOE Office of Science-supported Flash Center for 
Computational Science at the University of Chicago and the University of Rochester.
The authors acknowledge the OSIRIS Consortium, consisting of UCLA and IST (Portugal) for the use of the OSIRIS 4.0 framework.
This research used resources of the National Energy Research Scientific Computing Center (NERSC), a U.S. Department of Energy 
Office of Science User Facility located at Lawrence Berkeley National Laboratory, operated under Contract No. DE-AC02-05CH11231 
using NERSC award m1157. 
The authors would like to thank the OMEGA operations team for supporting this experiment
as well as R. Frankle and E. Doeg for processing the CR-39.
The authors also thank W. Fox, D. Schaeffer, and A. Milder for helpful discussions.

\bibliographystyle{aip.bst}
\bibliography{bibliography.bib}

\end{document}